\documentclass[final,5p,times,twocolumn]{elsarticle}

\usepackage{amssymb}
\usepackage{amsmath}
\usepackage{braket}
\usepackage{bbold}
\usepackage{hyperref}
\usepackage{multirow}
\usepackage{array}
\usepackage{booktabs}

\renewcommand{\bf}{\boldsymbol}
\newcommand{\Tr}{\,\hbox{\rm Tr}}

\newcolumntype{T}{!{\vrule width 1pt}}

\def\MSbar{\overline{\rm MS\kern-0.5pt}\kern0.5pt}

\begin{document}

\begin{frontmatter}

\title{A precise study of the SU(3) Yang-Mills theory across the deconfinement transition}

\author[a,b]{Leonardo Giusti}
\author[a,b]{Mitsuaki Hirasawa}
\author[b]{Michele Pepe}
\author[a,b]{Luca Virz\`\i}

\affiliation[a]{organization={Department of Physics, University of Milano-Bicocca},
                addressline={Piazza della Scienza 3},
                postcode={20126},
                city={Milano},
                country={Italy}}
                
\affiliation[b]{organization={INFN, sezione di Milano - Bicocca},
                addressline={Piazza della Scienza 3},
                postcode={20126},
                city={Milano},
                country={Italy}}

\begin{abstract}
We perform a detailed computation of key quantities across the first-order deconfinement phase transition of the SU(3) Yang-Mills
theory. Specifically, we calculate the entropy density, $s(T_c)/T_c^3$, on both sides of the transition and determine the latent heat 
$h$. The calculations are carried out in the lattice regularization with the Wilson action, employing shifted boundary conditions in the
temporal direction. Our simulations are performed at five different values of the lattice spacing in order to extrapolate the results to
the continuum limit. The latent heat can be measured also as the discontinuity in the trace anomaly of the energy-momentum
tensor: our result using the entropy density is compatible with the one obtained from the trace anomaly, giving a combined estimate
$h=1.175(10)$. Additionally, we determine the critical temperature $T_c$ in physical units with permille accuracy, yielding 
$T_c \sqrt{t_0} = 0.24915(29)$. These results allow to connect with precision the confined and the deconfined phases and we present an
improved computation of the Equation of State across the deconfinement transition for $T$ between 0 and $3.4 T_c$.
\end{abstract}

%\begin{keyword}
%Here
%\end{keyword}

\end{frontmatter}

%% \linenumbers

%% main text

\section{\label{sec:intro}Introduction}

Understanding the thermal properties of Quantum Chromodynamics (QCD) requires to delve into the dynamics of its gauge sector described by
the SU(3) Yang-Mills theory. As the temperature $T$ increases above a few hundreds MeV, pions cease to be approximate Goldstone bosons of
chiral symmetry and quarks pick up a mass of the order of the temperature. In this regime, as well as for higher temperatures, quarks are
heavy fields and the light modes are only gluons.

At variance with QCD that smoothly connects low and high temperatures, SU(3) Yang-Mills theory is characterized by a deconfinement phase
transition. While the first-order nature of that transition is well established, a precise quantitative
description is still lacking.  In fact, the strength of a first-order phase transition is quantified by the latent heat, $h$, which is given
by the discontinuity either in the energy density or, equivalently, in the entropy density at the critical temperature $T_c$. This is the
energy that is required to melt glueballs into a plasma of gluons. Despite many measurements of $h$ have 
been attempted (see~\cite{CELIK1983323,Svetitsky:1983bq,Karsch:1983ag,PhysRevLett.61.2058} for early studies), state of the art results are
still affected by large statistical and systematic errors and, moreover, they do not give a consistent
picture~\cite{Meyer:2009tq,Borsanyi:2012ve,Shirogane:2016zbf,Shirogane:2020muc,Borsanyi:2022xml}. 

At a first-order phase transition, the correlation length stays finite and there are typically coexisting phases with different energy
densities. This leads to metastability and hysteresis which may lock the system into a local minimum of one phase even
if the other phase would be energetically more convenient. Thus, a proper sampling of the coexisting phases requires overcoming an energy barrier by
tunnelling events whose probability is exponentially suppressed with the spatial size of the system. This makes the investigation by
numerical methods particularly challenging and, moreover, Monte Carlo algorithms -- except for a few cases -- are local and numerical
simulations performed in the critical region have long autocorrelation times.    

A standard approach to measure the thermal properties of the SU(3) Yang-Mills theory is based on computing the trace anomaly of the
energy-momentum tensor $T_{\mu\nu}$. This quantity is affected by an ultraviolet divergence, which comes from the mixing of
$T_{\mu\nu}$ with the identity operator and which must be subtracted. However, that ultraviolet divergence cancels out in the computation of
the latent heat since it can be evaluated as the difference in the trace anomaly between the confined and the deconfined phases at the
critical temperature.

The goal of this letter is to study the SU(3) Yang-Mills theory across the deconfinement phase transition with high accuracy. We first
compute the critical temperature $T_c$ with a 1\textperthousand\ precision in units of the reference scale
$\sqrt{t_0}$~\cite{Luscher:2010iy}, and the entropy density at the two sides of the phase transition, thus providing a determination of the
latent heat. We compute $h$ also from the discontinuity of the trace anomaly so to evaluate the same quantity using two observables with
different renormalization factors. The accurate determination of the latent heat then allows a precise computation of the Equation of State 
across the deconfinement phase transition. Using the data for the critical couplings and the determination of $T_c$ in units of
$\sqrt{t_0}$, we evaluate the entropy density, the pressure and the energy density for temperatures between 0 and $3.4 T_c$. This
investigation improves and complements the determination of the Equation of State carried out in Ref.~\cite{Giusti:2016iqr}.  

In this study we have considered the framework of shifted boundary conditions~\cite{Giusti:2011kt,Giusti:2010bb,Giusti:2012yj}
that represents a convenient setup for investigating the thermal properties of gauge theories by Monte Carlo
simulations~\cite{Giusti:2014ila,Giusti:2016iqr} since it avoids the issues related to the subtraction of ultraviolet divergences.
The use of an efficient method to determine the gauge coupling at the critical temperature has also been instrumental for this
study~\cite{Francis:2015lha}. 

The letter is organized as follows. In the next section, we outline the general framework of our investigation, along with the relevant
formulas used to compute the entropy density and the latent heat. In section~\ref{sec:betac}, we discuss the determination of $T_c$ and
of the critical couplings of the deconfinement phase transition in SU(3) Yang-Mills theory with shifted boundary conditions. Those values
are used in the next two sections: first, we present the results for the entropy density and the latent heat, then we update our estimate of
the Equation of State. Finally, in the last section, we summarize our findings and present our conclusions.

\section{\label{sec:setup}The general setup}
We formulate the SU(3) Yang-Mills theory on a (3+1)-dimensional lattice with size $L^3\times L_0$ and spacing $a$. The gauge action is
given by the Wilson plaquette action 
\begin{equation}\label{eq:action}
S[U] = \beta\! \sum_{x}\sum_{\mu<\nu} \left[ 1-\frac{1}{3} \mbox{Re} \Tr [U_{\mu\nu}(x)] \right],
\end{equation}
where the trace is over the color index and $\beta=6/g_0^2$ with $g_0^2$ being the bare gauge coupling. The plaquette field,
$U_{\mu\nu}(x)$, is given by the ordered product of the gauge field $U_\mu(x)\in$ SU(3) along the simplest non-trivial closed loop on the lattice  
\begin{equation}\label{eq:plaquette}
U_{\mu\nu}(x) = U_\mu(x)U_\nu(x+\hat\mu)U^\dagger_\mu(x+\hat\nu) U^\dagger_\nu(x).
\end{equation}
The gauge field is periodic in space while it has shifted boundary condition in the compact direction
\begin{equation}\label{eq:shiftedBC}
U_\mu(L_0,{\bf x}) = U_\mu(0,{\bf x}- L_0\, {\bf \xi})\; ,
\end{equation}
where the spatial vector $L_0\, {\bf \xi}$ has integer components in lattice units.
In the thermodynamic limit, this system~\cite{Giusti:2012yj} is equivalent to a periodic one at the temperature
$T^{-1}=L_0 \sqrt{\strut{1+{\bf \xi}^2}}$. 

In the framework of shifted boundary conditions, the off-diagonal elements of the energy-momentum tensor are no longer all vanishing and, in
particular, the entropy density $s(T)$ at the temperature $T$ is given by
\begin{equation}\label{eq:sT0k}
\frac{s(T)}{T^3} =  -\frac{(1+{\bf \xi}^2)}{\xi_k} \,
 \frac{Z_T \langle T_{0k} \rangle }{T^4}\; ,
\end{equation}
where $T_{0k}$ are the space-time components of the energy-momentum tensor on the lattice defined as
follows~\cite{Caracciolo:1989pt}  
\begin{equation}\label{eq:Tmunu}
T_{\mu\nu} =  \frac{1}{g_0^2}\Big\{F^a_{\mu\alpha}F^a_{\nu\alpha}
- \frac{1}{4} \delta_{\mu\nu} F^a_{\alpha\beta}F^a_{\alpha\beta} \Big\}\; .
\end{equation}
The field strength tensor $F_{\mu\nu}(x)=F^a_{\mu\nu}(x) T^a$ on the lattice is given by
\begin{equation}\label{eq:Fmunu}
F^a_{\mu\nu}(x) = - \frac{i}{4 a^2} 
\Tr\Big\{\Big[Q_{\mu\nu}(x) - Q_{\nu\mu}(x)\Big]T^a\Big\}\; ,
\end{equation}
with $T^a$ being the generators of the group SU(3) normalized as $ 2 \Tr [T^a T^b]=\delta_{ab}$.
The clover field $Q_{\mu\nu}(x)$ is defined as the sum of the 4 co-planar plaquettes resting on the point $x$
\begin{equation}\label{eq:Qmunu}
Q_{\mu\nu}(x) = U_{\mu\nu}(x) + U_{\nu-\mu}(x) + U_{-\mu-\nu}(x) +U_{-\nu\mu}(x)\; ,
\end{equation}
with the minus sign standing for the negative direction.

The lattice regularization breaks explicitly the invariance of the theory under translations and rotations with the consequence that the
energy-momentum tensor requires a finite multiplicative renormalization factor and $Z_T (g_0^2)$ is the one of the sextet
component. This quantity is part of the definition itself of $T_{\mu\nu}$ on the lattice and it has been computed non-perturbatively
in~\cite{Giusti:2015daa,Giusti:2016iqr}.  
 
At the critical temperature $T_c$, the confined and the deconfined phases coexist with two different values of the entropy density,
$s(T_c^-)$ and $s(T_c^+)$, respectively. These quantities can be measured separately by Monte Carlo simulations
performed at the critical temperature: the method consists in preparing the system in an initial state belonging to one of the two phases
and then running the numerical simulation on a system with a very large spatial size so to prevent tunneling events to the other phase. 
The latent heat is then readily defined as
\begin{equation}\label{eq:latheat}
  h = \frac{\Delta s(T_c)}{T_c^3} = \frac {s(T_c^+) - s(T_c^-)}{T_c^3}.
\end{equation}
An alternative way to compute $h$ consists in measuring the gap of the trace anomaly $A(T)$ of the energy-momentum tensor at the two
sides of the phase transition. One may use Eq.~(\ref{eq:Tmunu}) or else consider the definition related to the action
density~\cite{Boyd:1996bx}
\begin{equation}\label{eq:latheat_anomaly}
  h = \frac{\Delta A(T_c)}{T_c^4} = \frac{d\beta}{d\log(a)}\, \frac{a^4}{\beta L_0 L^3}\,
  \frac{\langle S(T_c^+) \rangle - \langle S(T_c^-) \rangle}{T_c^4}.
\end{equation}
An accurate determination of the dependence of the lattice spacing on the bare gauge coupling can be found in~\cite{Giusti:2018cmp}.

\section{\label{sec:betac} Determination of the critical coupling}
We have performed Monte Carlo simulations at the critical temperature $T_c$ for 5 values of the lattice spacing corresponding to systems with
temporal extension $L_0/a=5,6,7,8$ and 9 in lattice units and shift vector $\bf \xi = (1,0,0)$. Various observables can be
considered to define $\beta_c$, with different choices being equivalent in the thermodynamic limit. In our study, we
have used the quantity proposed in~\cite{Borgs:1991,Francis:2015lha}, which shows a rapid convergence to $\beta_c$ as the spatial volume
increases to the thermodynamic limit. This feature is particularly advantageous for investigating first-order phase transitions, where the
accurate determination of the critical coupling requires a proper sampling of the coexisting phases, achievable through frequent tunneling
events. However, since the probability of tunneling events is exponentially suppressed with increasing spatial volume, it is convenient to
avoid being constrained to large volumes. Otherwise, one would need to perform long Monte Carlo runs to capture a sufficient number of
tunneling events with a significant  increase of the computational effort.  

At the deconfinement phase transition the $\mathbb{Z} (3)$ center symmetry gets spontaneously broken: in the confined, unbroken phase we have
a single vacuum while in the deconfined, broken phase there are 3 degenerate vacua. The Polyakov loop is an order parameter for the breaking
of this symmetry and it is a useful quantity for characterizing the two phases. When shifted boundary conditions are taken into account, the
usual definition needs to be modified with an additional term given by a product of spatial links in order to close the path and make it
gauge invariant. The important issue is that the center symmetry $\mathbb{Z} (3)$ is unaffected by shifted boundary conditions and the
modified Polyakov loop is charged under that symmetry. In this study we consider the shift vector ${\bf \xi}=(1,0,0)$ since we
observed~\cite{Giusti:2016iqr} that lattice artefacts in $\langle T_{01} \rangle$ turn out particularly small. We thus define the Polyakov
loop $\Phi (\bf x)$ as follows   
\begin{equation}\label{eq:poly}
\Phi (\bf x)= 
\prod_{n=0}^{L_0/a-1} U_0(n a,\bf x) \prod_{n=0}^{L_0/a-1} U_1(0, \bf x _n )
\end{equation}
where $\bf x _n=\bf x-(L_0 -n a)\, \bf \xi $. It is useful to consider also the $\mathbb{Z} (3)$ projection 
\begin{equation} 
  \phi = \mbox{Re} \left\{ \frac {1}{(L/a)^3} \left( \sum_{\bf x} \frac{1}{3} \Tr [ \Phi (\bf x)] \right) \cdot \bar z \right\}
\end{equation}
with $z$ being the SU(3)  center element closest to the spatial average of $\Tr [ \Phi (\bf x)] /3$.

In the critical region, the probability distribution of $\phi$ exhibits two peaks corresponding to the confined and to the deconfined phases,
separated by a minimum located at $\phi_0$. In a finite volume, the identification of a field configuration as confined or deconfined is
somehow conventional: indicating with $\omega_{\rm c}$ and $\omega_{\rm d}$ the probability of the confined and of the deconfined phases,
respectively, following~\cite{Francis:2015lha} we define 
\begin{equation}
  \omega_{\rm c} (\beta,L/a) = \langle \theta ( \phi_0 - \phi ) \rangle; \quad
   \omega_{\rm d} (\beta,L/a) = \langle \theta ( \phi-\phi_0 ) \rangle
\end{equation}
where $\theta$ is the Heaviside step function. In~\cite{Borgs:1991,Francis:2015lha} the phase transition is defined to occur when 
\begin{equation}\label{eq:crit}
  \omega_{\rm d} = 3 \omega_{\rm c}
\end{equation}
where the factor 3 on the r.h.s. comes from the 3-fold degeneracy of the deconfined phase.
One can then consider the quantity 
\begin{equation}\label{eq:d}
  d(\beta,L/a) = \frac{ 3 \omega_{\rm c} (\beta,L/a) - \omega_{\rm d} (\beta,L/a)}{3 \omega_{\rm c} (\beta,L/a) + \omega_{\rm d} (\beta,L/a)}
\end{equation}
and the critical coupling is obtained as $\beta_c(L_0/a)=\lim_{L/a\rightarrow\infty} \beta_c(L_0/a,L/a)$, where $\beta_c(L_0/a,L/a)$ is
defined by the condition that $d(\beta_c(L_0/a,L/a),L/a)=0$.   

One can show that the above definition of $\beta_c$ is the same as the one coming from the peak of the Polyakov loop susceptibility in the
thermodynamic limit. In fact, as the spatial volume increases, the two peaks of the probability distribution of $\phi$ become narrow and
well-separated, allowing an easy identification of a field configuration as confined or deconfined.
In this simplified picture, the Polyakov loop susceptibility is proportional to $\omega_c \omega_d$ and, thus, it reaches its maximum when
$\omega_{\rm c} = \omega_{\rm d}$. Since both $\omega_{\rm c}$ and $\omega_{\rm d}$ scale exponentially with the spatial volume, the
difference from the condition in  Eq.~(\ref{eq:crit}) is power suppressed with the volume and vanishes in the thermodynamic limit. This
different finite-size scaling behavior leads to a faster convergence of $\beta_c(L_0/a, L/a)$ to its infinite-volume
limit~\cite{Borgs:1991} as defined in Eq.~(\ref{eq:d}) compared to the $(L/a)^{-3}$ approach based on the Polyakov loop susceptibility.   

\begin{figure}[h]
\centering
\includegraphics[width=.45\textwidth]{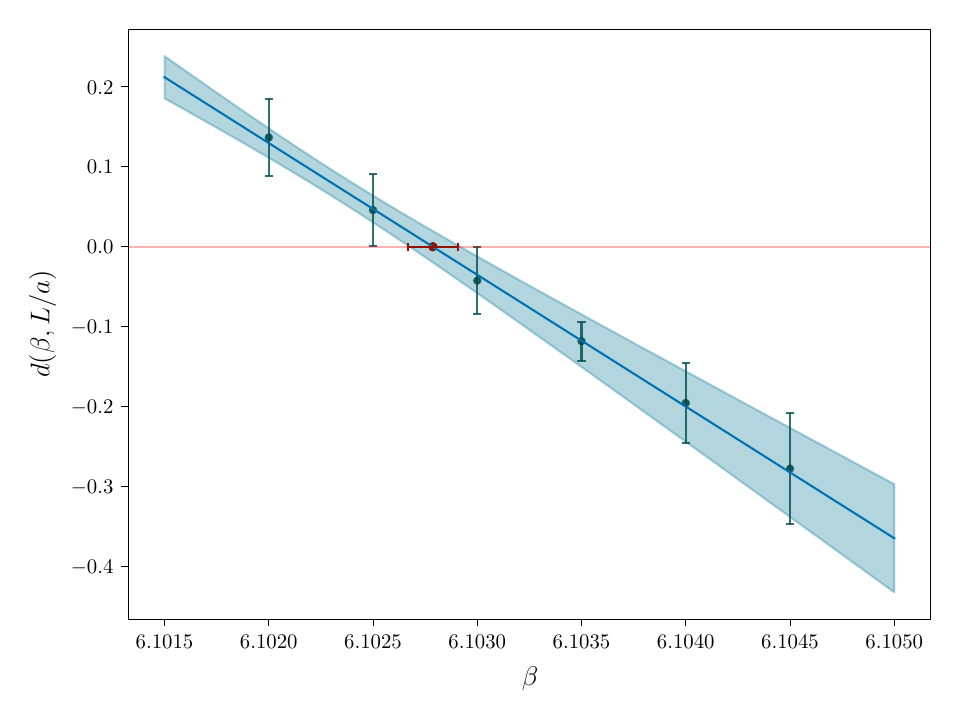}
\caption{Dependence of $d(\beta,L/a)$ on $\beta$ for $L_0/a=6$ and $L/a=48$. The band represents a fit of the data.\label{fig:beta_c}}
\end{figure}

\begin{figure}[h]
\centering
\includegraphics[width=.45\textwidth]{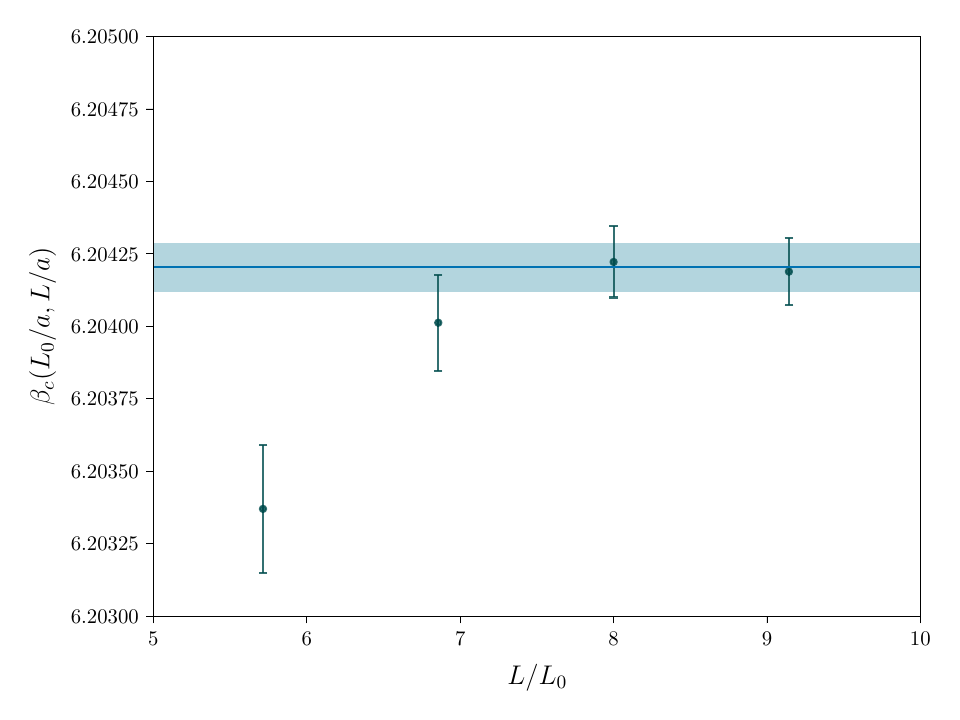}
\caption{Dependence of $\beta_c(L_0/a,L/a)$ on $L/L_0$ for $L_0/a=7$. The band represents the estimate of $\beta_c(L_0/a )$
  coming from the weighted average of the data obtained at the two largest values of $L/L_0$.\label{fig:beta_c_TL}} 
\end{figure}

In Figure~\ref{fig:beta_c} we show $d(\beta,L/a)$ as a function of $\beta$ for $L_0/a=6$ and $L/a=48$: we observe that the numerical data are
well-described by a line around $d=0$ and we estimate $\beta_c(L_0/a,L/a)$ by a linear fit of the data. 
A similar study is carried out for various values of $L/a$ at fixed $L_0/a$, with $L/L_0$ in the range between 5 and 12, so to be able to estimate
the critical coupling in the thermodynamic limit. Figure~\ref{fig:beta_c_TL} presents our results for $\beta_c(L_0/a,L/a)$ plotted as a
function of $L/L_0$, where  $L_0/a = 7$. The data show a rapid convergence to the infinite spatial volume limit which we estimate from a
weighted average of the data corresponding to the two largest investigated values of $L/L_0$.
Using this procedure we have computed the critical coupling for various values of $L_0/a$ with high accuracy and our results are reported in
Table~\ref{tab2:data}.

\begin{figure}[h]
\centering
\includegraphics[width=.45\textwidth]{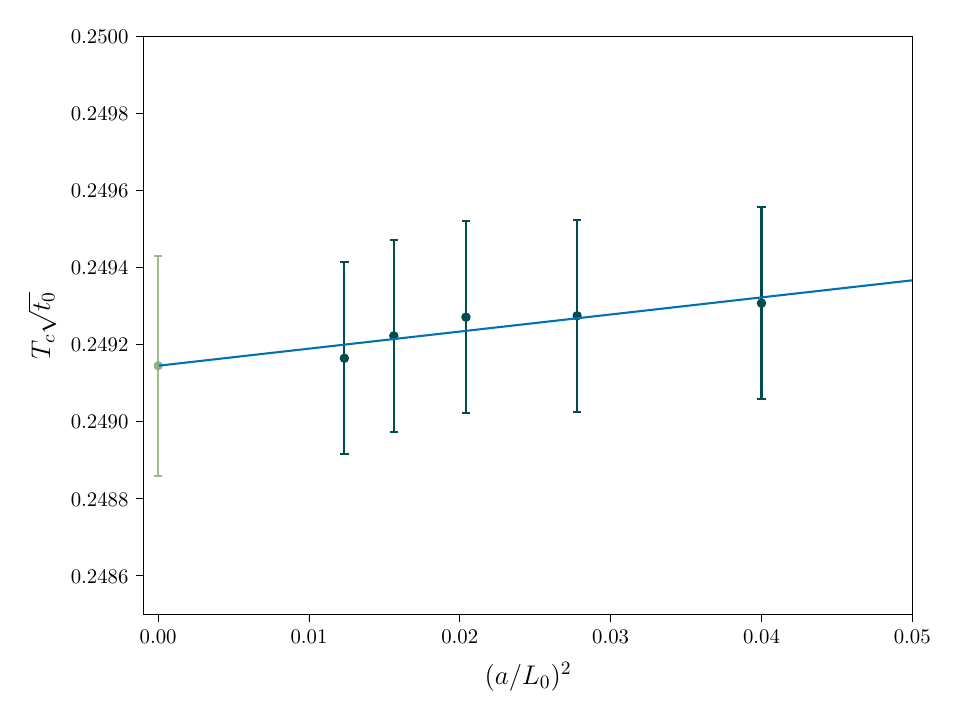}
\caption{Extrapolation to the continuum limit of $T_c \sqrt{t_0}$; the line represents a linear fit of the data.\label{fig:Tc}} 
\end{figure}

The computed values of the critical coupling allow us to express the critical temperature $a T_c= (a/L_0)/\!\sqrt{2}$ in physical units.
We consider the gradient-flow time $t_0$~\cite{Luscher:2010iy} to set the scale and we refer to the results in~\cite{Giusti:2018cmp} to
relate $t_0/a^2$ to the gauge coupling $\beta$. Lattice artefacts on $T_c \sqrt{t_0}$ are practically negligible as the data plotted in
Figure~\ref{fig:Tc} show; we consider a linear fit in $(a/L_0)^2$ and the extrapolation to the continuum gives
\begin{equation}\label{eq:Tc}
  T_c \sqrt{t_0} = 0.24915(29).
\end{equation}
with a final precision of 1\textperthousand\ mainly due to the accuracy with which the relation between $t_0$ and $\beta$ is known. Our
result is in agreement with the one computed in~\cite{Borsanyi:2022xml} in units of the scale $w_0$ once the relation with $\sqrt{t_0}$ in
taken into account~\cite{Knechtli:2017xgy}. 

\section{\label{sec:LH} Measurement of the latent heat $h$}
We compute the latent heat of the SU(3) Yang-Mills theory from the continuum limit extrapolation of the difference both of the entropy
density and of the trace anomaly between the deconfined and the confined phases, as discussed in section~\ref{sec:setup}. At criticality, in
the thermodynamic limit, the system remains locked in one phase without flipping to the other one. By considering very large spatial sizes
to make the probability of tunneling events negligible, we can unambiguously perform Monte Carlo simulations in either the confined or the
deconfined phase. The selection of the two phases is achieved at the beginning of the simulation by starting from an ordered configuration,
$U_\mu(x)=1\!\! 1$, for the deconfined phase and from a configuration thermalized at a temperature lower than $T_c$ for the confined one.   

In order to extrapolate the numerical results to the continuum limit, we have considered 5 values of the temporal extent, $L_0/a=5, 6,7,8$,
and 9, for which we have computed the critical coupling reported in Table~\ref{tab2:data}. Simulations with $L_0/a=5$ and 7 were performed
on lattices with spatial size $L/a=280$ while for $L_0/a=6,8$ and 9 we set $L/a=288$. These choices have been made to satisfy
the condition $L/(2 L_0)\in \mathbb{Z}$~\cite{Giusti:2012yj}, as required for $\bf \xi = (1,0,0)$, to avoid additional finite size effects
arising from the shifted boundary conditions. Furthermore, finite volumes effects are negligible compared to the statistical uncertainty
given the large spatial volumes used, as discussed in~Ref.~\cite{Giusti:2016iqr}.

\begin{table*}[th]
  \centering\small 
    \begin{tabular}{||c||c|c|c|c||c|c|c||c||}\hline
      $L_0/a$
      & $\langle T_{01}^- \rangle \times  10^6$ & $\mbox{Re} \langle \Tr [U_{\mu\nu}^-] \rangle/3  $
      & $s(T_c^-)/T_c^3$ & $n_{meas}$
      & $\langle T_{01}^+ \rangle \times  10^6$ & $\mbox{Re} \langle \Tr [U_{\mu\nu}^+] \rangle/3  $
      & $s(T_c^+)/T_c^3$ & $n_{meas}$   \\ \hline  
        5 & -5.898(20)  & 0.59272515(10)  & 0.2878(22) & $3.7\cdot 10^{5}$ & -31.90(8)  & 0.5928688(5)   & 1.556(12) & $4\cdot 10^{5}$ \\
        6 & -2.872(14)  & 0.60442466(5)   & 0.2894(24) & $5.5\cdot 10^{5}$ & -15.11(4)  & 0.60448650(25) & 1.523(11) & $7\cdot 10^{5}$ \\
        7 & -1.565(10)  & 0.61402046(5)   & 0.2916(26) & $7\cdot 10^{5}$   & -8.104(29) & 0.61405165(14) & 1.510(11) & $8\cdot 10^{5}$ \\   
        8 & -0.915(9)   & 0.62210597(4)   & 0.2908(34) & $4.5\cdot 10^{5}$ & -4.732(24) & 0.62212347(10) & 1.504(12) & $9\cdot 10^{5}$ \\
        9 & -0.566(10)  & 0.629047195(25) & 0.288(5)   & $10^{6}$          & -2.965(38) & 0.62905791(20) & 1.510(21) & $10^{6}$        \\ \hline
    \end{tabular}
    \caption{Expectation values of the bare matrix elements of the space-time components of the energy-momentum tensor $\langle
      T_{01}\rangle $ and of the plaquette $\mbox{Re} \langle \Tr [U_{\mu\nu}] \rangle/3$ in the confined (-) and in the deconfined (+)
      phases computed at $\beta_c(L_0/a )$. The data on the entropy density are obtained from Eqs.~(\ref{eq:sT0k})
      and using~Eq.~(\ref{eq:ZT}).}  
    \label{tab1:data}
\end{table*}

For this computation, as well as for the simulations performed to determine the critical couplings, we employed the standard over-relaxed
Cabibbo-Marinari algorithm~\cite{Creutz:1980zw,Adler:1981sn,Cabibbo:1982zn,Whitmer:1984he,Creutz:1987xi}. As expected, the Monte Carlo
simulations performed at $T_c$ have long autocorrelation times, on the order of several thousands sweeps, requiring large statistics to
obtain accurate numerical estimates. In Table~\ref{tab1:data} we report the collected statistics and the expectation values of the
space-time component of the bare energy-momentum tensor, $\langle T_{01} \rangle$, and of the average plaquette,
$\mbox{Re} \langle \Tr [U_{\mu\nu}] \rangle/3$, computed in the confined and in the deconfined phases at the values of $L_0/a$ that we
have considered. Using Eqs.~(\ref{eq:sT0k}) and~(\ref{eq:poly}) along with the finite renormalization constant $Z_T (g^2_0)$ computed
non-perturbatively in~\cite{Giusti:2015daa} and updated in~\cite{Giusti:2016iqr}  
\begin{equation}\label{eq:ZT}
Z_{_T}(g_0^2) = \frac{1 - 0.4367\, g_0^2}{1 - 0.7074\, g_0^2} - 0.0971\, g_0^4
 +  0.0886\, g_0^6 - 0.2909\, g_0^8.
\end{equation}
we have obtained the data reported in Table~\ref{tab1:data} for the entropy density at the critical temperature. The value of $g_0^2$ for
$L_0/a=5$ is by 1.5\textperthousand\ outside the range $[0,1]$ of Eq.~(\ref{eq:ZT}) and we slightly increased the uncertainty on
$Z_{_T}$ to take that into account.   

\begin{table}[th]
  \centering\small 
  \renewcommand{\arraystretch}{1.4}
    \begin{tabular}{||c||c||c|c||}\hline
      $L_0/a$ & $\beta_{\rm c} (L_0/a)$ & $ \Delta s(T_c)/T_c $  & $ \Delta A(T_c)/T_c $\\ \hline 
        5 & 5.99115(4)   & 1.269(10) & 1.266(9) \\
        6 & 6.10285(8)   & 1.233(10) & 1.227(6) \\
        7 & 6.20420(8)  & 1.218(10) & 1.215(6) \\   
        8 & 6.29626(12)  & 1.213(11) & 1.211(8) \\
        9 & 6.38017(16)  & 1.222(21) & 1.220(23)\\ \hline
    \end{tabular}
    \caption{Values of the critical couplings $\beta_c(L_0/a )$ in the infinite spatial volume limit using the shift vector ${\bf
        \xi}=(1,0,0)$ and expectation values of the latent heat as obtained using Eq.~(\ref{eq:sT0k}) (left column) or
      Eq.~(\ref{eq:latheat_anomaly}) (right column).}  
    \label{tab2:data}
  \renewcommand{\arraystretch}{1}
\end{table}

\begin{figure}[h]
\centering  
\includegraphics[width=.45\textwidth]{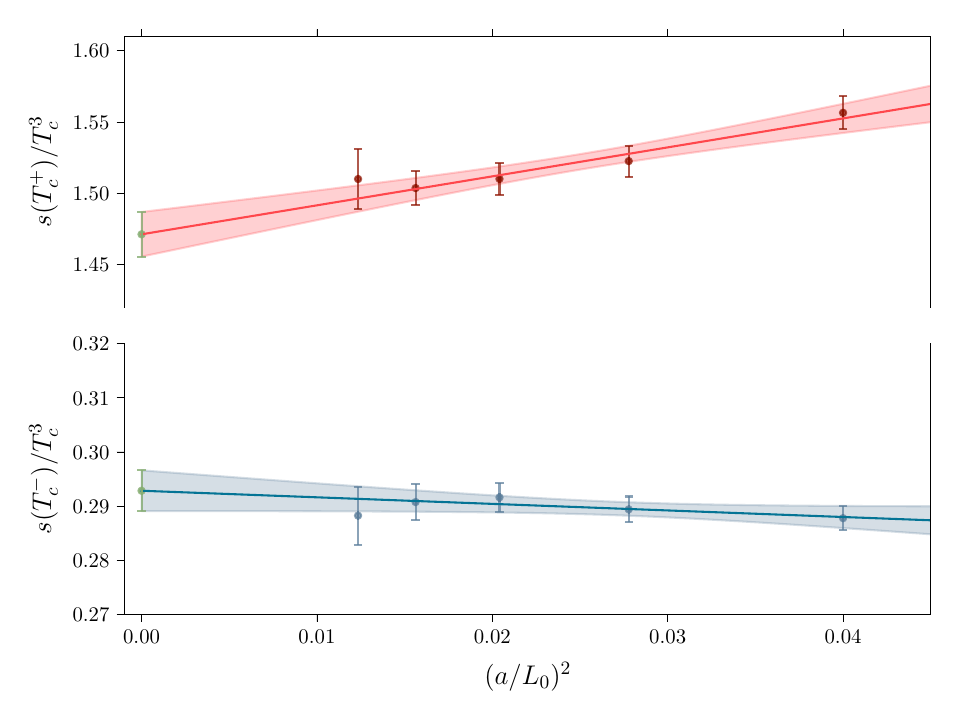}
\caption{Extrapolation to the continuum limit of the entropy density: the upper panel refers to the deconfined phase, $s(T_c^+)/T_c^3$, and
  the lower panel to the confined one, $s(T_c^-)/T_c^3$. The bands represent linear fits of the data.\label{fig:sT3_cold_hot}} 
\end{figure}

\begin{figure}[h]
\centering  
\includegraphics[width=.45\textwidth]{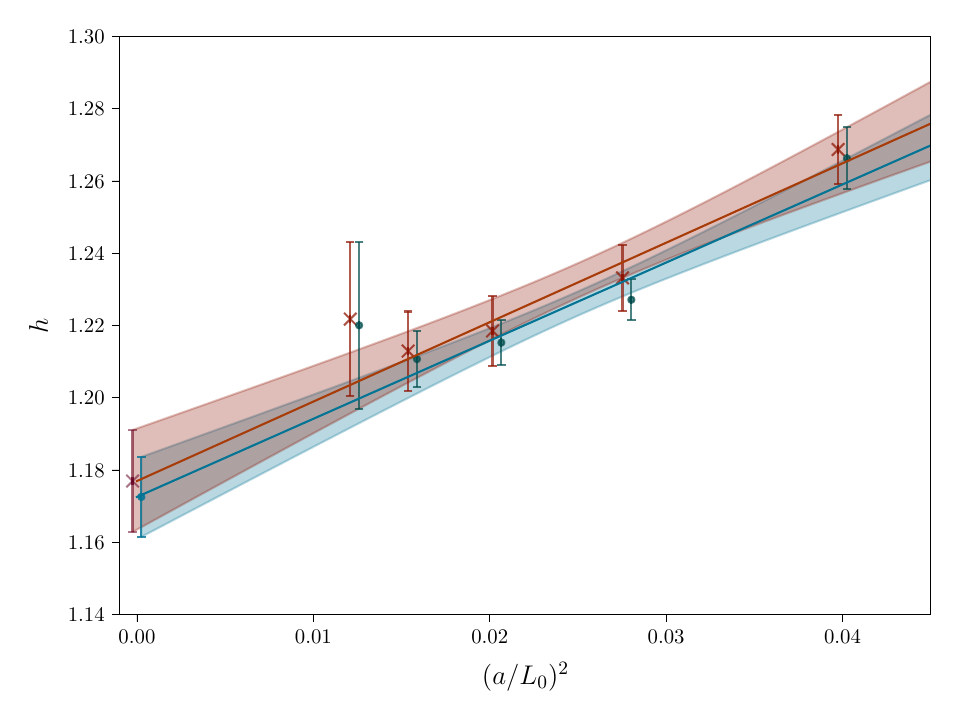}
\caption{Extrapolation to the continuum limit of the latent heat $h$. The crosses and the points are, respectively, the data obtained
  using the discontinuity in the entropy density and in the trace anomaly at $T_c$. The bands represent linear fits of the data. The data have been
  slightly displaced forward (points) and backward (crosses) to improve readability. \label{fig:latheat}}
\end{figure}

We can now extrapolate our results to the continuum limit: Figure~\ref{fig:sT3_cold_hot} displays the dependence on $(a/L_0)^2$ of the entropy
density in the confined (lower panel) and in the deconfined (upper panel) phases. Lattice artefacts are small in the confined 
phase and moderate in the deconfined one: in both cases a linear fit, represented by the blue band in the plots, provides a good
description giving the following results in the continuum limit 
\begin{equation}
  \frac{s(T_c^-)}{T_c^3}=0.2928(38); \quad \quad
  \frac{s(T_c^+)}{T_c^3}= 1.471(16).
\end{equation}
We measure the latent heat as the discontinuity in the entropy density or in the trace anomaly: these are two different observables, each
one with it own renormalization factor. In this way we can perform a consistency check of our measurement of $h$ and we report our numerical
results in Table~\ref{tab2:data}. In Figure~\ref{fig:latheat} we show the extrapolation to the continuum limit where one can observe that
lattice artifacts are moderate and follow a linear dependence on $(a/L_0)^2$. The best estimates we obtain for $h$ are
\begin{equation}
 h= \frac{\Delta s(T_c)}{T_c^3}=1.177(14); \quad \quad
 h= \frac{\Delta A (T_c)}{T_c^3}= 1.173(11).
\end{equation}
giving the combined result $h=1.175(10)$.

\section{\label{sec:EoS}The Equation of State}

We have used the determination of the critical couplings $\beta_c(L_0/a )$ listed in Table~\ref{tab2:data} to obtain the dependence of
$(a\,T_c)$ on the bare coupling. We have interpolated $\log (a\,T_c)$ with a polynomial in $(\beta-6)$ in the range $5.991 \leq \beta \leq 6.380$
\begin{equation}
  \log (a\,T_c) = \sum_{k=0}^3 c_k\, (\beta-6)^k
\end{equation}
with $c_0=1.62444$, $c_1=1.688386$, $c_2=-0.6487505$, and $c_3=0.4493516$. The fit gives $(a\,T_c)$ with an accuracy of about
$0.2 \text{\textperthousand}$. Using the above formula, we have determined the bare couplings to compute the entropy density for
temperatures between $1.01\, T_c$ and $1.10\, T_c$ at the 4 values of the lattice spacing corresponding to $L_0/a=5,6,7$, and 8 and shift vector
${\bf \xi}=(1,0,0)$. Then, considering the same temporal extensions in lattice units, we have used Eq.~(\ref{eq:Tc}) and the parametrization for
$t_0$ given in~\cite{Giusti:2018cmp} to find the gauge couplings corresponding to temperatures in the ranges $[0.80\, T_c, 0.98\, T_c]$
and $[1.15\, T_c, 2.5\, T_c]$. Also in these cases the shift has been set to ${\bf \xi}=(1,0,0)$.
In this way we fix the bare couplings with better accuracy than in Ref.~\cite{Giusti:2016iqr} using $r_0/a$~\cite{Necco:2001xg}. The
extrapolations to the continuum limit for the entropy density tend to compare better with the data of Ref.~\cite{Borsanyi:2012ve}
w.r.t. the results in Ref.~\cite{Giusti:2016iqr}.

Monte Carlo simulations have been carried out on lattices with spatial size $L/a=280$ for $L_0/a=5$ and 7 and with $L/a=288$ for $L_0/a=6$ and 8.
In Table~\ref{tab3:sT3} we report the computed values of $\langle T_{01} \rangle$ and the extrapolations of the entropy
density to the continuum limit obtained with linear fits in $(a/L_0)^2$. We note that, although we increased the uncertainty on $Z_{_T}$ for
$L_0/a=6$ at $0.80 T_c$ since the value of $g_0^2$ is by 5\textperthousand\ outside the applicability range $[0,1]$ of Eq.~(\ref{eq:ZT}), the accuracy of
the entropy density remains dominated by the error on $\langle T_{01} \rangle$ at this low temperature.
For the temperatures $T/T_c=$ 2.15, 2.30, and 2.50 the matrix
elements of the energy-momentum tensor have been improved at tree-level:
$\langle T_{01} \rangle \rightarrow \langle T_{01} \rangle \cdot (s_{cont}/s_{latt}(a/L_0, {\bf \xi}))$, where $s_{cont}/T^3=32\pi^2/45$ is
the entropy density in the Stefan-Boltzmann limit in the continuum and $s_{latt}(a/L_0, {\bf \xi})/T^3$ is the corresponding quantity
computed at tree-level in perturbation theory on the lattice with temporal size $L_0/a$ in lattice units and shift ${\bf \xi}$.
Using these new results, we obtain a better parametrization of the Equation of State for temperatures between
0 and $3.433 T_c$ and we can describe more accurately w.r.t. Ref.~\cite{Giusti:2016iqr} the change of the thermodynamic quantities as the
phase transition is crossed.  
\begin{table*}[th]
  \centering\small 
    \begin{tabular}{|cTccTccTccTccTcc|}\hline
      $\frac{L_0}{a}$ & $\beta$ & $\langle T_{01} \rangle \times  10^{5}$ &
      $\beta$ & $\langle T_{01} \rangle \times  10^{5}$ &
      $\beta$ & $\langle T_{01} \rangle \times  10^{5}$ &
      $\beta$ & $\langle T_{01} \rangle \times  10^{5}$ &
      $\beta$ & $\langle T_{01} \rangle \times  10^{5}$    \\

\specialrule{0.1em}{0em}{0em}
\hline
& \multicolumn{2}{cT}{$T= 0.80\, T_c$} 
& \multicolumn{2}{cT}{$T= 0.90\, T_c$} 
& \multicolumn{2}{cT}{$T= 0.95\, T_c$} 
& \multicolumn{2}{cT}{$T= 0.98\, T_c$} 
& \multicolumn{2}{c|}{$T= 1.01\, T_c$} \\

\hline

5  &    --   &     --      &   --    &    --       & 5.96102 & -0.2933(33) & 5.97897 & -0.4153(36) & 5.99701   &   -4.365(5)     \\
6  & 5.96703 & -0.0390(23) & 6.03683 & -0.0869(22) & 6.07017 & -0.1416(12) & 6.08970 & -0.2002(24) & 6.10923   &   -2.0756(16)     \\
7  & 6.05916 & -0.0206(17) & 6.13396 & -0.0462(15) & 6.16950 & -0.0773(13) & 6.19026 & -0.1066(13) & 6.21089   &   -1.1192(20)   \\
8  & 6.14424 & -0.0104(15) & 6.22300 & -0.0256(10) & 6.26021 & -0.0444(10) & 6.28188 & -0.0633(9)  & 6.30330   &   -0.6550(19)   \\

\hline

& \multicolumn{2}{cT}{$s(T)/T^3= 0.030(9)$} 
& \multicolumn{2}{cT}{$s(T)/T^3= 0.075(7)$} 
& \multicolumn{2}{cT}{$s(T)/T^3= 0.1419(37)$} 
& \multicolumn{2}{cT}{$s(T)/T^3= 0.198(4) $} 
& \multicolumn{2}{c|}{$s(T)/T^3= 2.046(19)$} \\

\specialrule{0.1em}{0em}{0em}
\hline

& \multicolumn{2}{cT}{$T= 1.02\, T_c$} 
& \multicolumn{2}{cT}{$T= 1.03\, T_c$} 
& \multicolumn{2}{cT}{$T= 1.04\, T_c$} 
& \multicolumn{2}{cT}{$T= 1.05\, T_c$} 
& \multicolumn{2}{c|}{$T= 1.06\, T_c$} \\
\hline 

5  &  6.00285   &   -5.004(7)    &  6.00865   &   -5.498(8)    &  6.01442   &   -5.907(9)     &  6.02017   &   -6.228(9) &  6.02588   &   -6.521(9)     \\
6  &  6.11555   &   -2.402(6)    &  6.12183   &   -2.624(8)    &  6.12807   &   -2.811(8)     &  6.13428   &   -2.988(8) &  6.14045   &   -3.119(8)     \\ 
7  &  6.21758   &   -1.284(5)    &  6.22423   &   -1.407(5)    &  6.23083   &   -1.512(5)     &  6.23739   &   -1.601(5) &  6.24391   &   -1.682(4)      \\ 
8  &  6.31026   &   -0.750(4)    &  6.31716   &   -0.824(4)    &  6.32400   &   -0.8786(38)   &  6.33080   &   -0.933(4) &  6.33754   &   -0.9780(38)   \\ 

\hline 
& \multicolumn{2}{cT}{$s(T)/T^3= 2.347(25)$} 
& \multicolumn{2}{cT}{$s(T)/T^3= 2.571(26)$} 
& \multicolumn{2}{cT}{$s(T)/T^3= 2.741(28)$} 
& \multicolumn{2}{cT}{$s(T)/T^3= 2.922(29)$} 
& \multicolumn{2}{c|}{$s(T)/T^3= 3.072(30)$} \\

\specialrule{0.1em}{0em}{0em}
\hline 

& \multicolumn{2}{cT}{$T= 1.08\, T_c$} 
& \multicolumn{2}{cT}{$T= 1.10\, T_c$} 
& \multicolumn{2}{cT}{$T= 1.15\, T_c$} 
& \multicolumn{2}{cT}{$T= 1.20\, T_c$} 
& \multicolumn{2}{c|}{$T= 1.25\, T_c$} \\

\hline

5  &  6.03722   &  -7.037(9)    &  6.04844   &  -7.409(10)     &  6.07563   &  -8.169(9)      &  6.10253   &  -8.704(9)  &  6.12879   &  -9.167(8)   \\
6  &  6.15269   &  -3.350(8)    &  6.16479   &  -3.543(8)      &  6.19417   &  -3.908(8)      &  6.22300   &  -4.177(8)  &  6.25104   &  -4.382(7)   \\
7  &  6.25682   &  -1.801(5)    &  6.26956   &  -1.905(4)      &  6.30044   &  -2.110(5)      &  6.33060   &  -2.251(6)  &  6.35982   &  -2.365(4)   \\
8  &  6.35088   &  -1.0566(38)   &  6.36402   &  -1.1154(38)    &  6.39605   &  -1.2294(37)    &  6.42708   &  -1.3125(39)&  6.45705   &  -1.377(4)   \\

\hline 
& \multicolumn{2}{cT}{$s(T)/T^3= 3.301(31)$} 
& \multicolumn{2}{cT}{$s(T)/T^3= 3.499(33)$} 
& \multicolumn{2}{cT}{$s(T)/T^3= 3.874(34)$} 
& \multicolumn{2}{cT}{$s(T)/T^3= 4.146(36)$} 
& \multicolumn{2}{c|}{$s(T)/T^3= 4.345(37)$} \\

\specialrule{0.1em}{0em}{0em}
\hline 

& \multicolumn{2}{cT}{$T= 1.278\, T_c$} 
& \multicolumn{2}{cT}{$T= 1.30\, T_c$} 
& \multicolumn{2}{cT}{$T= 1.40\, T_c$} 
& \multicolumn{2}{cT}{$T= 1.50\, T_c$} 
& \multicolumn{2}{c|}{$T= 1.60\, T_c$} \\

\hline

5  &  6.14321   &  -9.349(8)      &  6.15442   &  -9.508(8)   &  6.20387   &  -10.062(8)    &  6.25104   &  -10.451(9)&  6.29605   &  -10.766(8)    \\
6  &  6.26640   &  -4.499(5)      &  6.27830   &  -4.575(7)   &  6.33060   &  -4.830(8)     &  6.38013   &  -5.013(7) &  6.42708   &  -5.163(7)     \\
7  &  6.37578   &  -2.421(4)      &  6.38813   &  -2.450(4)   &  6.44219   &  -2.590(6)     &  6.49312   &  -2.695(6) &  6.54118   &  -2.771(4)      \\
8  &  6.47339   &  -1.4127(38)    &  6.48602   &  -1.430(4)   &  6.54118   &  -1.5112(39)   &  6.59296   &  -1.578(4) &  6.64173   &  -1.6148(38)    \\

\hline 
& \multicolumn{2}{cT}{$s(T)/T^3= 4.476(37)$} 
& \multicolumn{2}{cT}{$s(T)/T^3= 4.516(38)$} 
& \multicolumn{2}{cT}{$s(T)/T^3= 4.781(38)$} 
& \multicolumn{2}{cT}{$s(T)/T^3= 5.014(39)$} 
& \multicolumn{2}{c|}{$s(T)/T^3= 5.134(38)$} \\

\specialrule{0.1em}{0em}{0em}
\hline 

& \multicolumn{2}{cT}{$T= 1.80\, T_c$} 
& \multicolumn{2}{cT}{$T= 2.00\, T_c$} 
& \multicolumn{2}{cT}{$T= 2.15\, T_c$} 
& \multicolumn{2}{cT}{$T= 2.30\, T_c$} 
& \multicolumn{2}{c|}{$T= 2.50\, T_c$} \\

\hline 

5  &  6.38013   &  -11.239(8)  &  6.45705   &  -11.571(9)   &  6.51060   &  -11.754(9)   &  6.56097   &  -11.894(9)&  6.62377   &  -12.039(8)   \\
6  &  6.51405   &  -5.389(7)   &  6.59296   &  -5.541(7)    &  6.64763   &  -5.608(8)    &  6.69896   &  -5.695(7) &  6.76291   &  -5.762(7)    \\
7  &  6.62980   &  -2.883(4)   &  6.70995   &  -2.973(6)    &  6.76547   &  -3.020(5)    &  6.81767   &  -3.048(5) &  6.88291   &  -3.092(6)    \\
8  &  6.73153   &  -1.683(4)   &  6.81283   &  -1.735(4)    &  6.86931   &  -1.758(5)    &  6.92263   &  -1.785(5) &  6.98972   &  -1.8035(27)  \\

\hline

& \multicolumn{2}{cT}{$s(T)/T^3= 5.364(37) $} 
& \multicolumn{2}{cT}{$s(T)/T^3= 5.561(36) $} 
& \multicolumn{2}{cT}{$s(T)/T^3= 5.673(37) $} 
& \multicolumn{2}{cT}{$s(T)/T^3= 5.773(36) $} 
& \multicolumn{2}{c|}{$s(T)/T^3= 5.865(32) $} \\

\hline
\end{tabular}
    \caption{Expectation values of the bare matrix elements $\langle T_{01}\rangle $ of the energy-momentum tensor computed at the coupling
      $\beta$. Each data set refers to the temperature indicated at the top of the cell; the values 
      of the entropy density are obtained from Eqs.~(\ref{eq:sT0k}) and then extrapolated linearly in $(a/L_0)^2$ to the continuum limit
      with the result given at the bottom of same cells. At $T/T_c=$ 2.15, 2.30, and 2.50 the data have been extrapolated by
      improving $\langle T_{01}\rangle $ at tree-level.}  
    \label{tab3:sT3}
\end{table*}

For temperatures $T/T_c \in [1,3.433]$ we consider a Pad\'e fit
\begin{equation}\label{eq:EoS_ent_aTc}
  \frac{s(T)}{T^3} = \frac{s_1+s_2\, w+s_3 \, w^2}{1+s_4\, w+s_5 \, w^2}
\end{equation}
where $w=\log(T/T_c)$ and $s_1=1.471$, $s_2=312.8$, $s_3=5155$, $s_4=141$, $s_5=865.5$ for  $T/T_c \in [1,1.1]$ while for  $T/T_c \in
[1.1,3.433]$ we have $s_1=1.742$, $s_2=64.21$, $s_3=138.2$, $s_4=15.08$, $s_5=18.26$. The uncertainty associated to this
formula is about 1\% up to $T/T_c =1.15$ and then decreasing linearly to 0.55\% at $T/T_c =2.5$ where it flattens.
Below $T_c$, the relative accuracy drops due to the fast decrease of the entropy density as $T$ goes to 0. At the lowest investigated
temperature, $T/T_c=0.8$, our data agrees with the expectation of a model based on a gas of non-interacting relativistic bosons
where the pressure given by a particle of mass $m$ at temperature $T$ is
\begin{equation}
  p(m,T) = k \frac{(m T)^2}{2 \pi^2} \sum_{n=1}^\infty \frac{1}{n^2} K_2 \left(n \frac{m}{T} \right) \,,
\end{equation}
with $k$ being the multiplicity of the state and $K_2$ the modified Bessel function. We have considered the lightest glueball states $0^{++}$,
$0^{-+}$, $2^{++}$, and $2^{-+}$ (the values of the glueball masses are taken from Table 34 in~\cite{Athenodorou:2021qvs}) with mass lower
than $2 m_{0^{++}}$; the multiplicity is $k=(2 J+1)$ in terms of the spin $J$. As the temperature increases towards $T_c$, 
heavier states become relevant and the Hagedorn spectrum~\cite{Hagedorn:1965st} describes how they contribute to the entropy density,
despite being suppressed due to their larger masses. We observed that a phenomenological fit $(a_0+a_1 t^{1/3}+a_2 t^{2/3}+a_3 t)$ with
$a_0=0.2927$, $a_1=0.0466$, $a_2=-1.9425$, $a_3=1.8595$ and $t=(1-T/T_c)$ provides a good interpolation of our numerical data of $s/T^3$ in the
range $T/T_c \in [0.8,1]$. The accuracy is 2-2.5\% in the range $[0.95,1]$, increasing to about 10\% when the temperature goes down
to $T/T_c =0.9$ and staying constant in absolute value to $T/T_c =0.8$. 

We can determine the pressure by integrating $s(T)$ in the temperature. In the range $T/T_c \in [0,0.8]$ we integrate the entropy density as
it results from the glueball gas model and then we consider a trapezoidal interpolation of $\log(s(T)/T^3)$ to integrate up to $T_c$. A cubic
function represents well the pressure in the range $[0.8,1]$ 
\begin{equation}\label{eq:EoS_p_lowT}
  \frac{p(T)}{T^4} = p_0+ p_1\, t +p_2\, t^2 +p_3 \, t^3
\end{equation}
where $p_0=0.01784$, $p_1=-0.19222$, $p_2=1.012867$, and $p_3=-2.14708$ with an accuracy that decreases linearly from 0.002 to 0.001.

Above $T_c$ we parametrize the pressure with a Pad\'e function 
\begin{equation}\label{eq:EoS_p}
  \frac{p(T)}{T^4} = \frac{p_1+p_2\, w+p_3 \, w^2}{1+p_4\, w+p_5 \, w^2},
\end{equation}
for  $T/T_c \in [1,1.1]$ we have $p_1=0.01768$, $p_2=1.846$, $p_3=50.27$, $p_4=17.53$, $p_5=6.008$, while for
$T/T_c \in [1.1,3.433]$ the parameters are $p_1=-0.009149$, $p_2=2.519$, $p_3=6.341$, $p_4=1.874$, $p_5=3.556$.
The accuracy increases linearly from 0.001 to 0.0085 in the range $[1,1.5]$ and then flattens.

\begin{figure}[h]
\centering
\includegraphics[width=.45\textwidth]{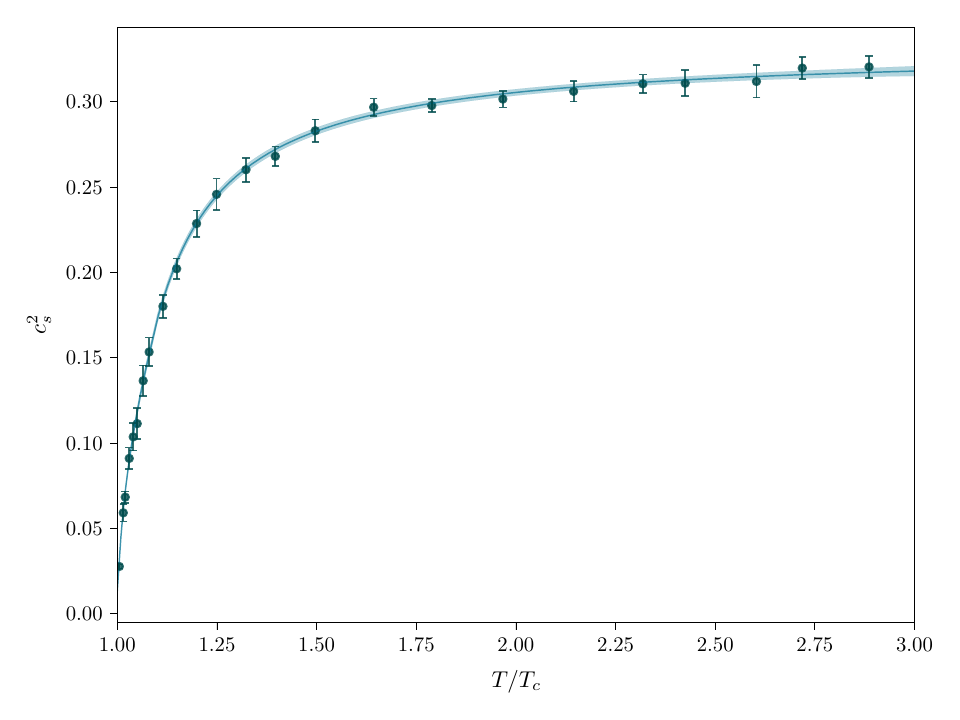}
\caption{Dependence of the speed of sound on the temperature in the deconfined phase. The blue band results from the Pad\'e fit of
  the Monte Carlo data of $s(T)/T^3$ above $T_c$.\label{fig:sound}}
\end{figure}

Finally, using the thermodynamic relation $e=T s - p $, we can derive the energy density: the absolute uncertainty on $e$ is the same as the
one on the entropy density up to $T\sim 1.2 T_c$ while at higher temperatures it decreases to be about 80\% of that error. We also notice that the
square of the speed of sound $c_s^2= \frac{dp}{de}=(\frac{d\log s}{d\log T})^{-1}$ can be computed from the entropy density and
we present it in Fig.~\ref{fig:sound}. This result can be similarly extended to higher temperatures using the data of
Ref.~\cite{Giusti:2016iqr}.

The above parametrizations represent our best estimates for the Equation of State of the SU(3) Yang-Mills theory across the deconfinement
phase transition; these results complement the one given in~\cite{Giusti:2016iqr} for higher temperatures.

\section{\label{sec:conc}Discussion and conclusions}

In this study we have investigated the thermal features of the SU(3) Yang-Mills theory across the deconfinement phase transition using
shifted boundary conditions where the primary observable is the entropy density. In this framework, we determined the 
critical temperature $T_c$ in physical units with a 1 permille precision as reported in Eq.~(\ref{eq:Tc}). We computed the entropy
density on both sides of the deconfinement phase transition thus measuring the latent heat $h$ and we then checked the compatibility of
this determination with the result obtained considering the discontinuity of the trace anomaly at $T_c$. Our final best estimate
$h=1.175(10)$ was obtained with 1\% accuracy by extrapolating the data of Monte Carlo simulations to the infinite spatial volume limit and to the
continuum limit. This represents a significant improvement compared to the current best estimates available in the
literature. In~\cite{Shirogane:2020muc}, the  computation is carried out for systems with the two fixed aspect ratios $L/L_0=6$ and 8. While
our result is compatible within just over one standard deviation with the value at $L/L_0=8$ computed with a 3.6\% accuracy, the large
systematic uncertainties prevent the authors of Ref.~\cite{Shirogane:2020muc} from obtaining a reliable estimate in the thermodynamic
limit. In Ref.~\cite{Borsanyi:2022xml} the latent heat has been computed in the continuum and in the thermodynamic limits with 2\%
statistical and  2.6\% systematic uncertainties. Combining those two errors either in quadrature or linearly, we observe a tension with our
result. Furthermore, the precise determination of the relation between the gauge coupling and the lattice spacing in
Ref.~\cite{Giusti:2018cmp} enabled us to obtain an accurate and complete description of the Equation of State of the SU(3) Yang-Mills theory
across the deconfinement phase transition. 
A precise determination of the Equation of State of Yang-Mills theory across $T_c$ can be also useful as a prototype for new gauge forces
that can be relevant in models of dark matter (see e.g.~\cite{Boddy:2014yra,Soni:2016gzf,Laine:2024wyv}). It provides insights into the thermodynamic
properties and the phase transitions of a strongly interacting gauge theory, which can influence the behavior and formation of dark matter
in the early universe.  

\textbf{Acknowledgments.}
We acknowledge YITP in Kyoto University for granting us access to the supercomputer Yukawa-21. We also thank CINECA for providing us with a
very generous access to Leonardo during the early phases of operations of the machine and for the computer time allocated thanks to
CINECA-INFN and CINECA-Bicocca agreements. The R\&D has been carried out on the PC clusters Wilson and Knuth at Milano-Bicocca. This work is
(partially) supported by ICSC - Centro Nazionale di Ricerca in High Performance Computing, Big Data and Quantum Computing, funded by
European Union - NextGenerationEU. We wish to thank Leonardo Cosmai for useful discussions and Isabella Leone Zimmel for her contribution
during the early stages of this collaboration.

%\bibliographystyle{elsarticle-num} 
%\bibliography{bibfile}

\end{document}